# Design and testing of high-speed interconnects for Superconducting multi-chip modules

S. Narayana[1,3], V. K. Semenov[1], Y. A. Polyakov[1], V. Dotsenko[2], and S.K. Tolpygo[1,2,*]

[1] Department of Physics and Astronomy, Stony Brook University, Stony Brook, NY 11794, USA
[2] HYPRES, Inc., 175 Clearbrook Road, Elmsford, NY 10523, USA
[3] Rowland Institute at Harvard, Harvard University, MA 02139, USA
[*] Present address: MIT Lincoln Laboratory, 244 Wood Street, Lexington, MA 02420, USA

**Abstract.** Superconducting single flux quantum (SFQ) circuits can process information at extremely high speeds, in the range of hundreds of GHz. SFQ circuits are based on Josephson junction cells for switching logic and ballistic transmission for transferring SFQ pulses. Multi-chip modules (MCM) are often used to implement larger complex designs, which cannot be fit onto a single chip. We have optimized the design of wideband interconnects for transferring signals and SFQ pulses between chips in flip-chip MCMs and evaluated the importance of several design parameters such as the geometry of bump pads on chips, length of passive micro-strip lines (MSL)s, number of corners in MSLs as well as flux trapping and fabrication effects on the operating margins of the MCMs. Several test circuits have been designed to evaluate the above mentioned features and fabricated in the framework of 4.5-kA/cm$^2$ HYPRES process. The MCMs bumps for electrical connections have been deposited using the wafer-level electroplating process. We have found that, at the optimized configuration, the maximum operating frequency of the MCM test circuit, a ring oscillator with chip-to-chip connections, approaches 100 GHz and is not noticeably affected by the presence of MCM interconnects, decreasing only about 3% with respect to the same circuit with no inter-chip connections.

## 1. Introduction

Very long ago it was recognized that superconducting transmission lines provide extremely low dispersion and, therefore, are suitable for transferring short pico-second pulses over relatively long distances [1] and for chip-to-chip communication [2]. Currently various kinds of superconducting transmission lines, e.g., coplanar lines (CL) [3], strip lines (SLs) and microstrip lines (MSLs) [4] are vitally important components of superconductor digital circuits. For instance, they are widely used for inter-chip routing of Single Flux Quantum (SFQ) pulses in integrated circuits (ICs).

Most of the requirements for transmission lines, such as low losses and low signal distortion, are identical in semiconductor and superconductor technologies. There is one difference, however. All digital semiconductor devices are suited for operation with high impedance loads. This enables using of transmission lines with the highest possible characteristic impedance, and 50-Ω transmission lines are used almost exclusively. In contrast, the natural impedance of transmission line in superconductor electronics is around 4 Ω as it provides the best matching to currently available Josephson junctions [5], [6]. With future technology advances [7], the matched impedance can increase up to 10 Ω – 15 Ω. Such low impedance lines are exotic for room-temperature electronics. They became useful only together with Josephson junctions. At optimal impedance matching [8] the known drivers and receivers required for connecting active Josephson circuitry by transmission lines contain only a few Josephson junctions, see, for example, [8], [9]. At such low junction overhead, superconducting transmission lines have already become routine intra-chip connections when miniaturization and circuit density are of primary concerns.

The required miniaturization is achieved by using thin-film microstrip lines (Fig. 1) in which the signal strip and the ground plane are located on the same side of the substrate. Characteristic impedance $\rho$ is inversely proportional to width of the signal strip $W$:

$$\rho \sim \sqrt{t/W}, \tag{1}$$

and can be easily adjusted to the desired level by a simple manipulation of the strip width. It is more difficult to manipulate by the thickness of the dielectric, $t$. However, integrated circuits have several (three to six) levels of metallization separated by dielectric films, and a desired dielectric thickness from ~0.1 μm up to ~1 μm could be selected simply by placing signal strips on different superconducting layers. High impedance (50 Ω) transmission lines are also important for superconductor electronics because they serve as components of interfaces between semiconductor and superconductor circuits. However, they are quite similar to well investigated semiconductor counterparts [10] and require no special attention here.

In recent years there has been significant interest in developing multi-chip module (MCM) technology for superconducting digital circuits in order to increase functionality and complexity of superconducting digital systems. This requires transferring SFQ (clock and data) pulses between chips with rates comparable to intra-chip rates in order to realize the full ultra-high speed potential of superconductor ICs. The most challenging components of such MCMs are inter-chip connections. The fact of successful transmission of SFQ pulses between chips was reported long time ago [11], [12]. Unfortunately all earlier reports [11]-[15] either did not share interconnect design details or presented results of numerical simulations without experimental results [15]. In all these earlier works, the inter-chip data transfer rates were found to be very sensitive to the size and geometry of solder bumps used for chip-to-chip connections.

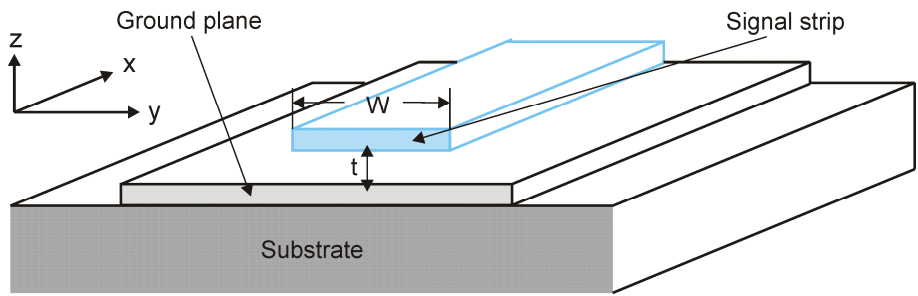

Figure 1. Sketch of a superconductor thin film microstrip line.

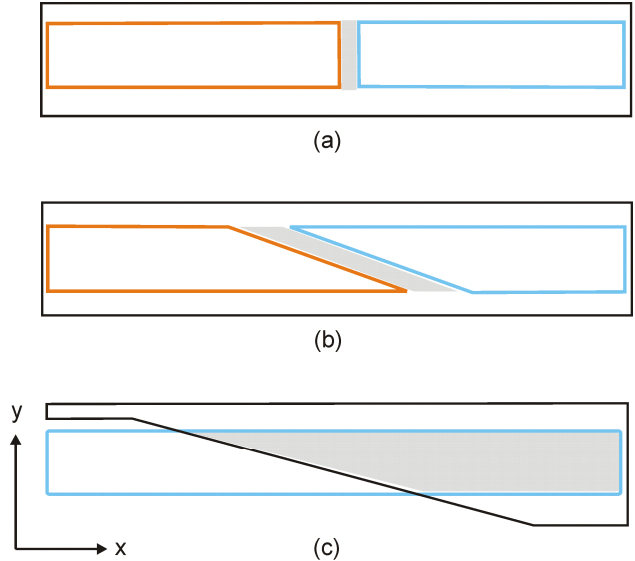

Figure 2. Straight (a) and tapered (b, c) cuts of microstrip lines is an essential part of the suggested bump design technique. Bonding stuff connecting the left and right strips in Fig. 1(a) and Fig. 1(b) is shown in grey.

Electrical connections for signal transfer (in z-direction) between two chips in a flip-chip configuration are provided via some conducting medium (solder bumps, balls, gold studs, etc.) connecting transmission lines on one chip with the corresponding lines on the other chip. This transition in z-direction should provide very good impedance matching in order to maximize the interconnect bandwidth and minimize signal reflection. Various highly developed interconnect technologies exist for MCMs with semiconductor ICs (see, e.g. [10]. So, it may seem to be a simple matter to adapt them to superconducting circuits. However, the mentioned above low $\rho$ value of superconducting transmission lines requires special optimization of interconnects that will be considered in this paper.

Our technique for connecting low impedance microstrip lines by evaporated gold bumps and some experimental results on data transfer rates in superconducting flip-chip MCMs were presented earlier [16]. Recently we have carried out more comprehensive optimization of wideband bumps and developed a more advanced fabrication technique with more compact bumps. Here we report our recent findings and, in particular, results of investigations of dependences of flip-chip interconnect performance on several factors including MSL length, the number and geometry of bumps connecting MSLs located on different chips, etc. We also show that our connection technique, while sensitive to thin film patters in the chip (x-y) plane, is quite insensitive to the bump geometry (bump shape and height). This insensitivity of the proposed technique to the features of chip connecting media (bumps) is a great advantage over the previously used approach [14] that, in contrast, is very sensitive to the details of bump geometry as discussed in the cited paper.

## 2. Design of chip interconnects

To introduce our technique for connecting low impedance microstrip lines located on different chips, let us consider two simplified examples. Assume we need to join together signal strips of two similar MSLs, see Fig. 2(a), using some bonding medium that is slightly unmatched, e.g., non-superconducting. (Let us repeat that this is the artificial example.) It is intuitively clear that the impact of mismatching would be weaker if the strip edges in the joint are not straight (Fig. 2a) but tapered as shown in Fig. 2(b). In this case the contact area is larger and undesirable contact resistance of the joint is lower. Since our real tapering technique is quite complicated, we think it is appropriate to show yet another example of tapering in Fig. 2(c). Here our goal is to taper the electromagnetic width (and hence the impedance) of a microstrip without tapering its signal strip. This can be achieved by tapering a much wider ground plane (contoured by black line in Fig. 2(c)) under the strip.

To connect two MSLs located on different chips, it is necessary to connect their signal strips and ground planes. In the flip-chip configuration, the assembly is much easier if geometries and vertical (layer) structure of layers to all bump connections on both chips are made identical (Fig. 3(a)), and only conducting joints (bumps) are placed between them. We strictly follow this recommendation, and our design technique is restricted to optimization of thin film patterns of interconnects in the vicinity of the bumps. Our goal is to make electrodynamics properties of the signal bump connection similar to the discussed earlier (Fig. 2(b)) properties of a tapered joint of two strips. Since the bump material is usually non-superconducting and we want to decrease both its resistance and the contact resistance of the joint, the bump diameter should usually be made larger than the width of the strips that is typically ~ 10 μm or less. This requires making a widening of the signal strip (a contact pad) for placing there the bump. An example of this widening, making a circular contact pad, is shown in Fig. 3(b). In this case, the current from the signal strip on the base chip (shown in blue) flows up (in $z$-direction) to the strip on the flip chip (orange) along the perimeter of the bump. Because of "impedance tapering" the current is uniformly distributed along the perimeter. The required tapering is achieved by making a narrow circular cut (a moat) in the ground plane film around the bump contact pad, shown as a white donut in Fig. 3(c), and a proper shift ("misalignment") of the circular signal pad with respect to the donut-shaped gap (moat) in the ground plane, as shown as dC in Fig. 3(c). The complementary signal bump contact pad on the flip chip has exactly the same design but is simply mirror-inverted with respect to the $y$-axis, as shown in Fig. 3(b). As a result, the whole structure after flip-chip bonding should behave almost as a uniform microstrip line which has a transition in the $z$-direction from the MCM carrier (base chip) onto the flipped chip. To reduce the resistance of the non-superconducting joint, the bump height should be made as low as technologically possible, just enough to provide for the needed compression.

A major deviation from this ideal model is caused by four ground bumps surrounding, but remote, from the signal bump and providing electrical connection of the ground planes of two chips. (In general, the ground bumps can be shared between the neighboring MSLs.) It is well known that in an MSL the signal current flows in the microstrip while the return current mainly flows in the ground plane area under the microstrip and the fringing is usually small. However, this is not the case in a flip-chip MCM because the return current must flow via the ground bumps in order to get from chip to chip. So the return current must "depart" from under the signal strip. Such distribution of the return current causes an additional parasitic inductance. This parasitic inductance is low because the magnetic field induced by this current is "boxed" in a small volume between two ground planes (the distance between the ground plane of two chips in $z$-direction is much less than the distance between bumps in $x$-$y$ plane). Since we deal with mainly 2D problem, the inductance

$$L_p = L_\square \cdot N \tag{2}$$

is measured in the number of squares that only logarithmically depends on the radius of contact pad, $R1$, and the distance between the signal and ground bumps, $R2$:

$$N = \ln[R2/(R2-R1)]/2\pi. \tag{3}$$

The sheet inductance $L_\square = \mu_0 \cdot t_m$ is defined by magnetic gap $t_m = t + 2\lambda$, where $\lambda$ is magnetic field penetration depth and $t$ is the spacing between the carrier and the flip chip ground planes. At the typical values, $t_m = 3$ µm and $L_\square \approx 3.8$ pH. At $R2/(R2-R1) = 5$ the number of squares equals 0.25. As a result, the parasitic inductance $L_p$ is about 1 pH. Thus, even at a low impedance of the MSL (e.g., $\rho \sim 4\ \Omega$), the cutoff frequency $f \sim \rho/2\pi L_p$ corresponding to this parasitic inductance is ~670 GHz. In practice, the cutoff frequency can be even higher. This is because the estimated lump inductance $L_p$ is only a simplified model for an inductance distributed along the microstrip line. The small impact of the parasitic inductance could be further compensated by a local increase of the signal strip widening.

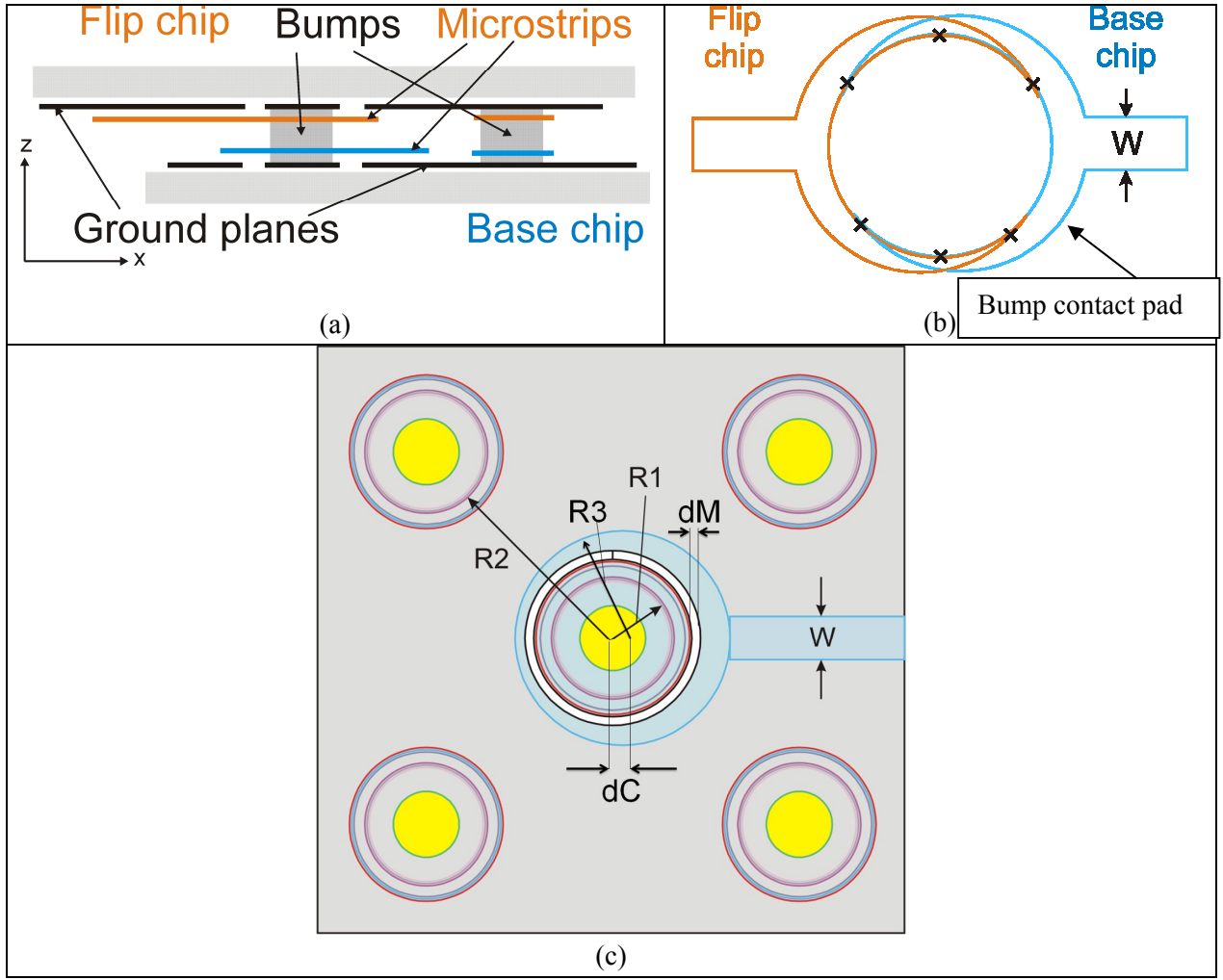

Figure 3. Bump connection of two microstrip lines in a flip-chip MCM. The signal and the ground plane contact pads for bumps have identical sequences of circuit layers and the same dimensions of all contact holes between them, and only differ by the patterns of signal and ground plane films. Electrically, the signal connection can be presented as an edge connection of two MSLs with varying widths (b). In the real connection (c), the signal bump in symmetrically surrounded by 4 ground plane bumps. The white ring around the signal contact pad shows a moat in the ground plane film isolating the signal bump from the ground plane. The bumps (shown in yellow) are formed on the contact pads at the ends of microstrips and on the ground planes of the base and flip chips. All bumps have the same diameter, smaller than the diameter of contact pads to allow for bump compression at flip-chip bonding. Here

dM is width of the ground plane moat, dC is "misalignment" between centers of the bump and circular widening of the signal strip, R1 is radius of bump, R2 is distance between the signal and ground bumps, R3 is the radius of the widening of the signal strip.

We complemented the described qualitative design considerations by numerical simulations using Full-wave 3D Planar Electromagnetic Field Solver Software [17]. During the simulations we indeed could observe some resonant reflection related to mismatch effect discussed in the previous paragraph. The reflection was observed to be either inductive or capacitive, depending on numerical value of R3 (see Fig. 3(c)) that served as the main adjustable parameter. The optimized value of R3 corresponds to no noticeable reflection in simulations. At this value of R3, an excessive capacitance of the signal contact pad circle exactly compensates the parasitic inductance.

Note that it is difficult to compare our approach with the one used in [14]. This is because we optimized the matching by tuning patterns of superconducting thin films deposited on the base and flip chips. In our approach, the matching conditions only weakly (logarithmically) depend on the dimensions and distances between the bump pads. In z-direction, our approach relies on the use of very short bumps giving small parasitic inductance. In contrast, the authors of Ref. [14], considering large bumps, illustrated that the matching depends mostly on dimensions and shape of solder bumps and distance between them, whereas the thin film patterning was not taken into the consideration. As a result, the floor plans are even not shown in the cited paper.

Similarly to the described MSL interconnects, we also optimized flip-chip interconnects between 50-Ω coplanar waveguide (CPW) lines serving for delivery of RF signals and clock to chips in MCMs. The proper matching was achieved by tapering (widening) the CPW gap in the ground plane and the central line of the CPW around the signal bump contact pad.

## 3. Fabrication of circuits and MCMs

Superconducting integrated circuits used in this work for MCM testing have been fabricated using HYPRES process for superconductor electronics and Nb/Al/AlO$_x$/Nb Josephson junctions with 4.5 kA/cm$^2$ critical current density [6], [7]. This process uses four Nb superconducting layers: 100-nm-thick ground plane layer, M0, and three wiring layers M1 (150 nm), M2 (300 nm), and M3 (500 nm). All contact pads for placing bumps present a stack containing all these layers, so the pad can be attached to any microstrip line formed in any of the superconducting layers. All dielectric layers in the stack are removed. The total thickness of Nb stack is, hence, 950 nm. The top surface of the stack was covered by a layer of underbump metallization (UBM) consisting of Mo/Pd/Au multilayer with the following thicknesses of the layers: 50 nm (Mo), 200 nm (Pd), and 100 nm (Au).

In our previous work, bumps for flip-chip bonding were formed at the very last step of the wafer fabrication by using a wafer-level lift-off process and e-beam evaporation of the bump material (Au or Cu with Au capping layer) [16]. In this work we have developed an electroplating process for depositing bumps in order to minimize the use of gold evaporation and the process cost.

For electroplating of uniform bumps all contact pads where bumps are intended to be placed need to be electrically connected. To achieve this we changed a few last steps in the standard HYPRES process [18] as follows. After the last Nb wiring layer deposition (layer M3), the entire wafer surface is covered by a well conducting metal. In the standard HYPRES process [18], the next step would be patterning of this layer that would create electrically isolated contact pads. Instead, we used this continuous M3 layer to perform electroplating. Firstly, a UBM was formed on M3 layer by a lift-off process and e-beam evaporation. A multilayer of Mo/Pd/Au with the thicknesses given above was used. The lift-off mask was formed by an image reversal process using AZ5214E photoresist. After the lift-off, a new photolithography was done to create circular openings in the photoresist for electroplating of bumps on the UBM. Since the rate of electroplating depends on the local current density, all bumps were chosen to have the same diameter (30 μm) in order to obtain uniform bump heights. Secondly, indium bump electroplating was done using an electrolytic cell with indium sulfamate bath [19] and a solid In anode. The total plating current was 20 mA, and the typical plating time was 3 min. The typical bump height was 5 μm. Indium bumps were plated on contact pads located on flip chips and on the carrier (base chip). The typical picture of the plated bump is shown in Fig. 4. After the plating, the photoresist mask was removed and the wafer was cleaned. Finally, the M3 layer was patterned by using photolithography and reactive ion etching. During the etching In bumps were protected by photoresist mask. The final view of the fabricated interconnects is shown in Figure 5 for a 50-Ω CPW line.

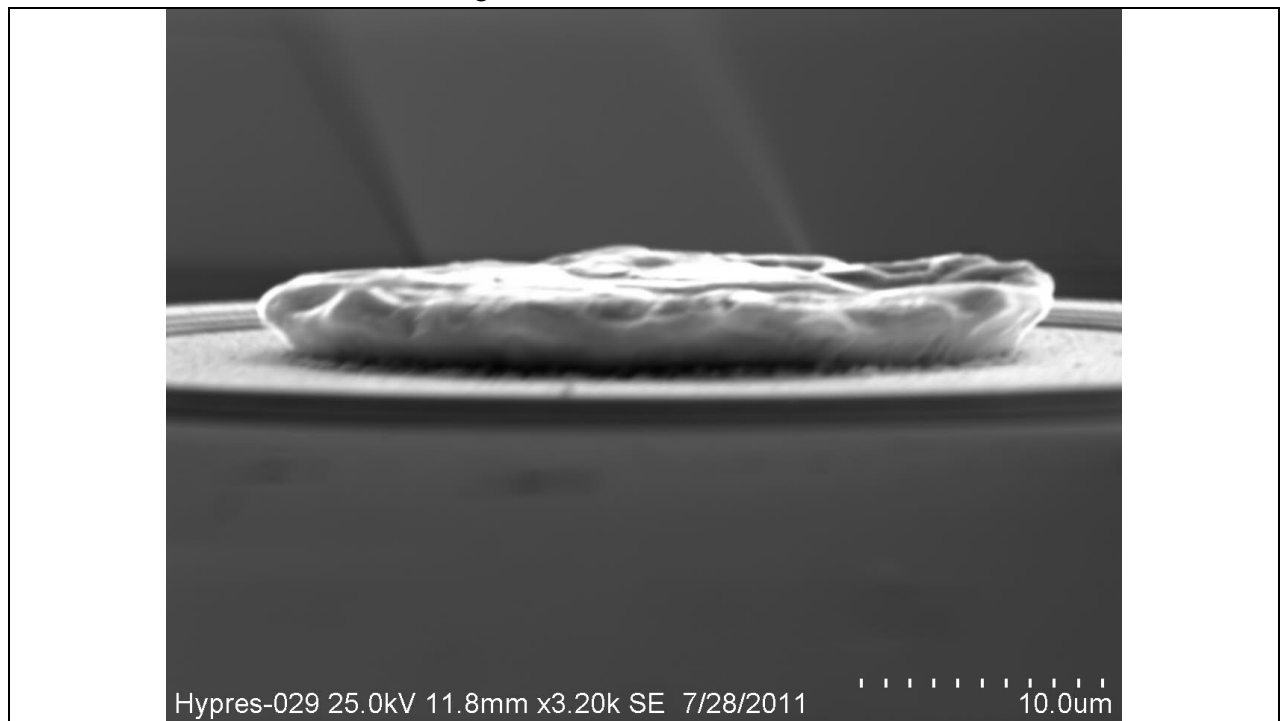

Figure 4. SEM picture of electroplated In bump on top of Nb contact pad with underbump metallization. Bump diameter is 30 μm and height is 4 μm. It has a slight mushroom shape due to slope of walls of plating holes in photoresist.

After wafer dicing, the MCMs were assembled on FC-150 flip chip bonder. A force of ~ 5 g per bump was applied to press In bumps on the flip chip into the corresponding bump on the carrier and join them together. A cryogenic adhesive (epoxy Tra-Bond 2115) was added between the chip and the carrier to increase the bonding strength, stability, and improve heat conduction between chips [20]. The adhesive was cured at 65 °C for 90 min. The adhesive was chosen to have the thermal expansion coefficient larger than that of Si and Nb. Therefore, the bumps become additionally compressed upon MCM cooling.

For many practical applications, MCMs would need to operate not immersed in liquid helium but in vacuum on closed cycle cryocoolers. This chip cryopackaging is dominant for the existing and planned superconductor electronics systems including Digital-RF systems described in [21]-[25]. Thermal conductivity of MCMs is then important because heat released on superconducting chips has to be removed via the MCM bumps and cryogenic adhesive to the cryocooler's cold plate without heat exchange liquid or gas. Thermal conductivity of the adhesive used in our flip-chip MCM at low temperatures was studied in [20]. It was found that heat conduction through the adhesive is larger than through the metal bumps. Additionally, the heat conduction in the MCM can be improved by loading the adhesive with carbon nanotubes [26].

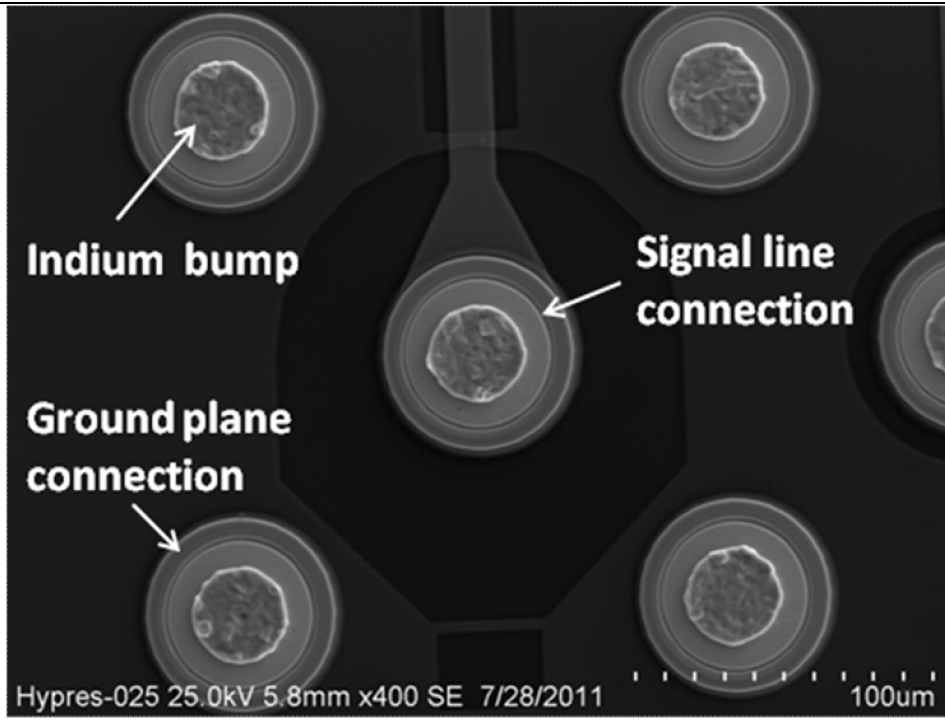

Figure 5. SEM picture of the fabricated CPW interconnects, showing electroplated In bumps in the centers of all contact pads. Each connection to a signal line (shown in the center) is surrounded by four ground plane connections which can be shared between adjacent signal lines. Niobium ground plane film is removed around the signal line connection and under the signal line to provide a proper 50 Ω impedance of the line and optimize the interconnect bandwidth. These areas look darker black in the SEM image due to fewer backscattered electrons.

## 5. MCM testing technique

Our approach to characterization of SFQ pulse transmission through inter-chip connections is based on a conventional ring oscillator shown in Fig. 6 and described, for example, in [4]. The ring contains a Josephson Transmission line (JTL) and a microstrip line that can be "broken" by one or several transitions through bumps. The rate of pulse transmission is controlled in two different ways. A coarse control is provided by the number of SFQ pulses circulating in the ring. The number of pulses can be changed by injecting a new pulse via a merger cell that is also part of the ring. A fine control is provided by adjusting the bias current applied to the JTL. The quality of the oscillator and, therefore, of the chip-to-chip connections is characterized by margins for the bias current applied to the receiver (marked as **Res** in Fig. 6) that converts electromagnetic pulses propagating along the MSL into SFQ pulses. The operation of the ring is monitored by measuring the dc voltage and, therefore, the rate of pulse travelling along the ring. (The measured dc voltage is related to the number and the oscillation frequency of SFQ pulses in the ring by the Josephson relation.) Transmission failures are detected via spontaneous changes of the voltage corresponding to a loss or creation of an extra pulse.

One reference circuit was intentionally laid out without chip-to-chip connections and used to evaluate the properties, in particular the maximum frequency of operation, of the circuit without any bump crossings. All test circuit components were designed using Josephson junction logic blocks developed at Stony Brook, and were designed to minimize parasitic flux trapping using methods described in [27, 28].

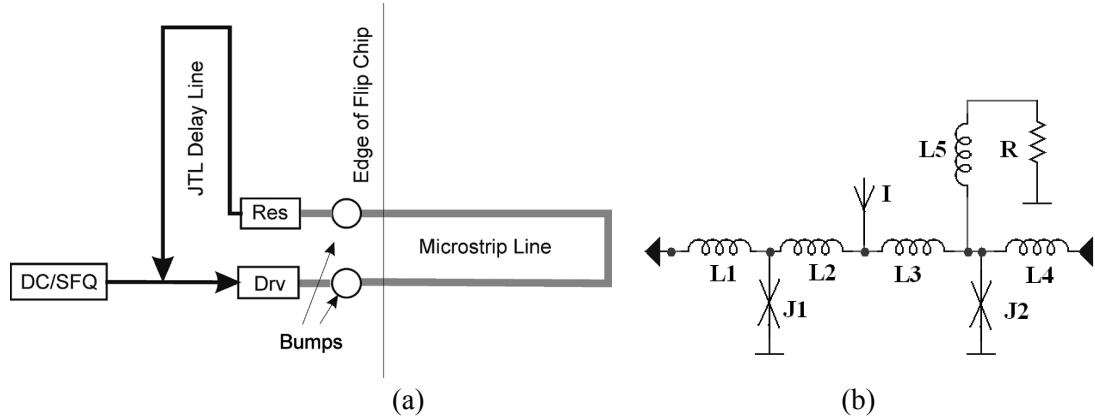

(a) (b)

Figure 6. (a) Block diagram of the ring oscillator used for testing inter-chip SFQ pulse transmission. SFQ driver and receiver are marked Drv and Res, respectively. (b) The element of the Josephson Transmission Delay Line (JTDL) with L5/R ~ 5 ps.

All JTL sections include LR components shown in Fig. 6(b). They create a strong anti bunching effect and therefore provide a more uniform distribution of pulses along the ring. The signal strips of MSLs were laid out in M2 layer, while M0 layer served as the ground plane. Lengths of the MSLs were varied from 1.8 mm to 20 mm. We used both straight MSL and meandered ones (containing 90° turns) to increase the length. Although a 90-degree turn of microstrip line usually does not cause any noticeable problems at microwave frequencies, it might be reasonable to consider such corners as a potentially disturbing factor for circuits operating with picosecond SFQ pulses having THz frequency components. All design variations and parameters of studied interconnects are given in Table 1 and Table 2. Test circuits 4, 6, 7, 9, and 10 each have modified signal bump pads and in circuits 4, 9, 10 we have 4 bumps interrupting the MSLs because we expect the adverse effects of bump connections to multiply. In test circuit 6, the width of the moat, dM, isolating the signal pad and the ground plane is increased from 5 μm to 10 μm. Test circuits 4, 9, and 10 have 4 bumps interrupting the MSL and vary in the values of dC and R3 which control the misalignment and overlapping of M2 structure in the signal pad as shown in Figure 3, whereas dM=5 μm is kept the same. In test circuit 7, the M2 misaligned structure was completely removed, that is expected to convert the signal bump into a low frequency structure.

## 6. Experimental Results

The measurements have been carried out in a cryo-cooler setup. To prevent the undesirable flux trapping we followed the thermo-cycling and magnetic shielding procedures described in [27],[28]. The measurement results were quite reproducible if we followed the shielding precautions.

Let us start with the reference test circuit laid out with MSL without bump crossings. The microstrip line is 20 μm wide, 1950 μm long, and contains four 90 degree turns. The measurement results are collected into two plots shown in Fig. 7. The left plot (Fig. 7(a)) shows the upped and lower margins for receiver bias current as a function of the pulse circulation rate. The plot shows the margins at 18 different circulation rates. In fact, we can extract margins at any particular rate because, as was mentioned earlier, the rate can be tuned by changing the JTL bias. Figure 7(b) shows these transmission rates at a different number (one to eighteen) of pulses travelling along the ring. It is easy to check that at higher rates the shown traces cover all frequencies. It means that it is possible to set SFQ pulse transmission at any rate within the covered range and measure the receiver margins at this particular rate. However, this procedure is quite time consuming. To speed up the measurements, all other circuits have been investigated at one JTL bias current corresponding to the center of its measured margins. According to Fig. 7(a) the margins are almost frequency independent at lower rates and partially degrade above ~ 80 GHz. Ninety six GHz was the highest frequency of stable operation for this reference circuit. At higher frequencies the margins sharply collapse to zero and no data are presented. The highest observed on-chip transmission frequency is somewhat below our expectations and the previously obtained results for circuits with evaporated bumps [16]. It is determined by the maximum speed of operation of circuits made in this particular fabrication run and has nothing to do with interconnects. However it is sufficiently high to investigate the disturbing impacts of interconnect matching and bump bonds on chip-to-chip transmission.

Test results for circuits with inter-chip connections are shown in Figs. 8-10 and summarized in Table 2. As was said, the reference circuit (Expt. 1) demonstrated the best performance with 96 GHz maximum frequency. The presence of any bump connections degrades the circuit performance (the maximum transmission rate). However, the observed degradation is quite small. For the optimized pad geometry (Expts. 2, 3, 5 and 8) the maximum frequency was in the range from 80 GHz to 87 GHz. The degradation is stronger for interconnects with intentionally distorted pad geometries. The worst performance with about 60 GHz maximum rate was observed in Expt. 4, where pads have been intentionally made without any misalignment between the pad and M2 disk (dC=0). Interconnects with a wider cut in the ground plane (Expt. 6) also show more degraded performance with maximum frequency about 67 GHz.

Table 1. Parameters of chip interconnects in different test circuits (ring oscillators) studied

| Test Circuit | Bump pad design | Comment |
| --- | --- | --- |
| Expts. 2, 3, 5, 8 | dM=5, R3=45, dC=7.5 | The best geometry according to numerical simulations. |
| Expt. 4 | dM=5, R3=45, dC =0 | No misalignment between the pad and M2 disk; |
| Expt. 6 | dM=10, R3=45, dC=7.5 | The width of the ground plane donut hole is increased. |
| Expt. 7 | dM=5, R3=45, dC=0 | No high-frequency tuning structure |
| Expt. 9 | dM=5, R3=50, dC=10 | The capacitance of M2 and the misalignment are increased. |
| Expt. 10 | dM=5, R3=40, dC=7.5 | The capacitance of M2 circle is reduced |

Table 2. Parameters of chip-to-chip connections and the maximum frequency of operation

| Circuit name | Length of MSL, μm | Number of bumps | Number of 90 degree turns | Maximum frequency, GHz |
| --- | --- | --- | --- | --- |
| Expt. 1 (Ref.) | 1,950 | 0 | 4 | 96 (MCM #1) |
| Expt. 2 | 4,100 | 2 | 4 | 80 (MCM#1) |
| Expt. 3 | 2,300 | 2 | 4 | 82 (MCM #1); 76 (MCM #2) |
| Expt. 4 | 2,450 | 4 | 10 | 60 (MCM #1) |
| Expt. 5 | 19,570 | 2 | 16 | 82 (MCM #1) |
| Expt. 6 | 2,150 | 2 | 4 | 67 (MCM #1) |
| Expt. 7 | 4,350 | 2 | 4 | None (MCM #1) |
| Expt. 8 | 4,350 | 2 | 4 | 87 (MCM #1) |
| Expt. 9 | 2,450 | 4 | 10 | 93 (MCM #1); 82 (MCM #2) |
| Expt. 10 | 2,450 | 4 | 10 | 72 (MCM #1) |

In Expt. 9 we intentionally decreased the capacitance of M2 disk by making it smaller with respect to the one optimized in electromagnetic simulations. This circuit demonstrated the best performance, giving the highest maximum frequency (~93 GHz) of chip-to-chip transmission among all studied interconnects. This experimentally found adjustment to the optimized in simulations geometry should be regarded as the experimentally optimized (new optimal) geometry. It is likely that our simulations slightly underestimate some parasitic, and thus a small reduction in the tuning capacitance improves the performance of bump-bonded interconnects. At this new optimum, the maximum chip-to-chip SFQ pulse transmission rate through 4 bump bonds in series is only ~ 3% lower than in the circuits without any bump crossings.

As expected, in a test circuit with removed impedance matching structure in the interconnect (Expt. 7), no chip-to-chip SFQ pulse transmission through the MSL interrupted by this interconnect was found. These experiments confirm the importance of a careful high frequency design of the chip-to-chip connections and high sensitivity of the maximum inter-chip transfer rates to relatively small variations in the interconnect tuning structure, which is not surprising given that pico-second SFQ pulses contain very high frequency components.

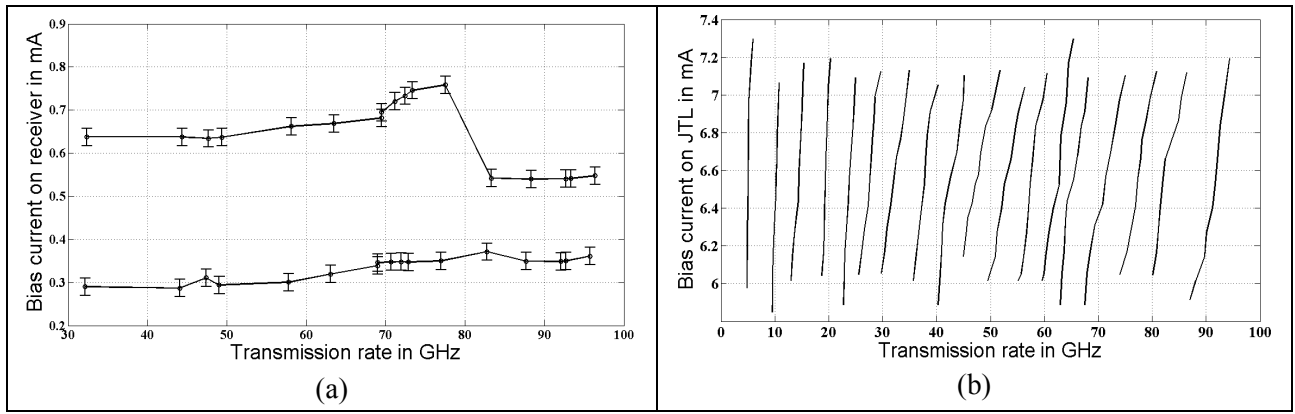

Figure 7. (a) The upper and lower margins of the receiver bias current as a function of SFQ pulse transmission rate for the reference circuit. (b) The frequency of transmission of SFQ pulses in the ring oscillator at various numbers of fluxes circulating in the ring for the typical MCM. The maximum of 18 pulses can be inserted into the ring oscillator. The speed of circulation of the pulses could be controlled by varying the bias on the JTL.

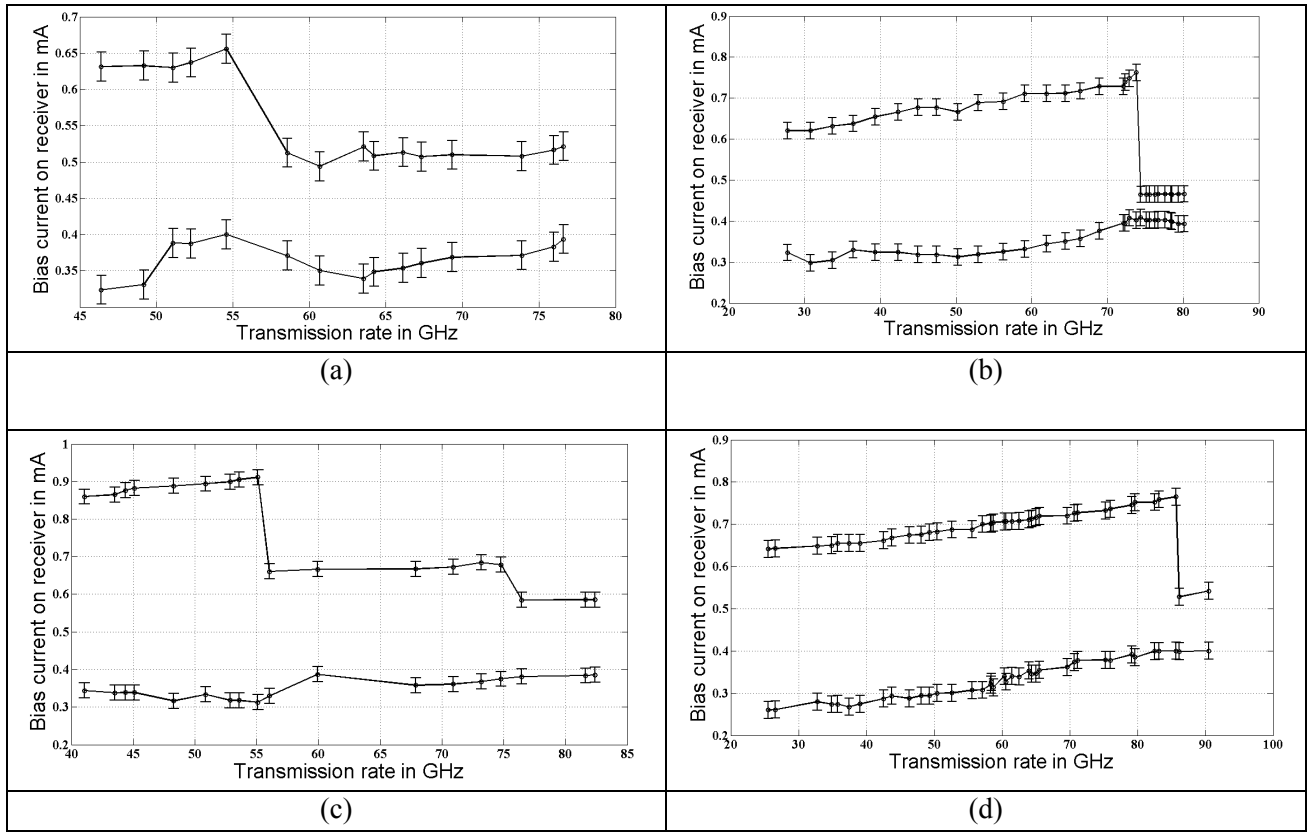

Figure 8. The upper and lower margins of the receiver bias current as a function of SFQ pulse transmission rate between the flip-chip and the MCM carrier for test circuits with interconnect design optimized in simulations (dM=5, R3=45, dC=7.5): (a) Expt. 2; (b) Expt. 3; (c) Expt. 5; (d) Expt. 8

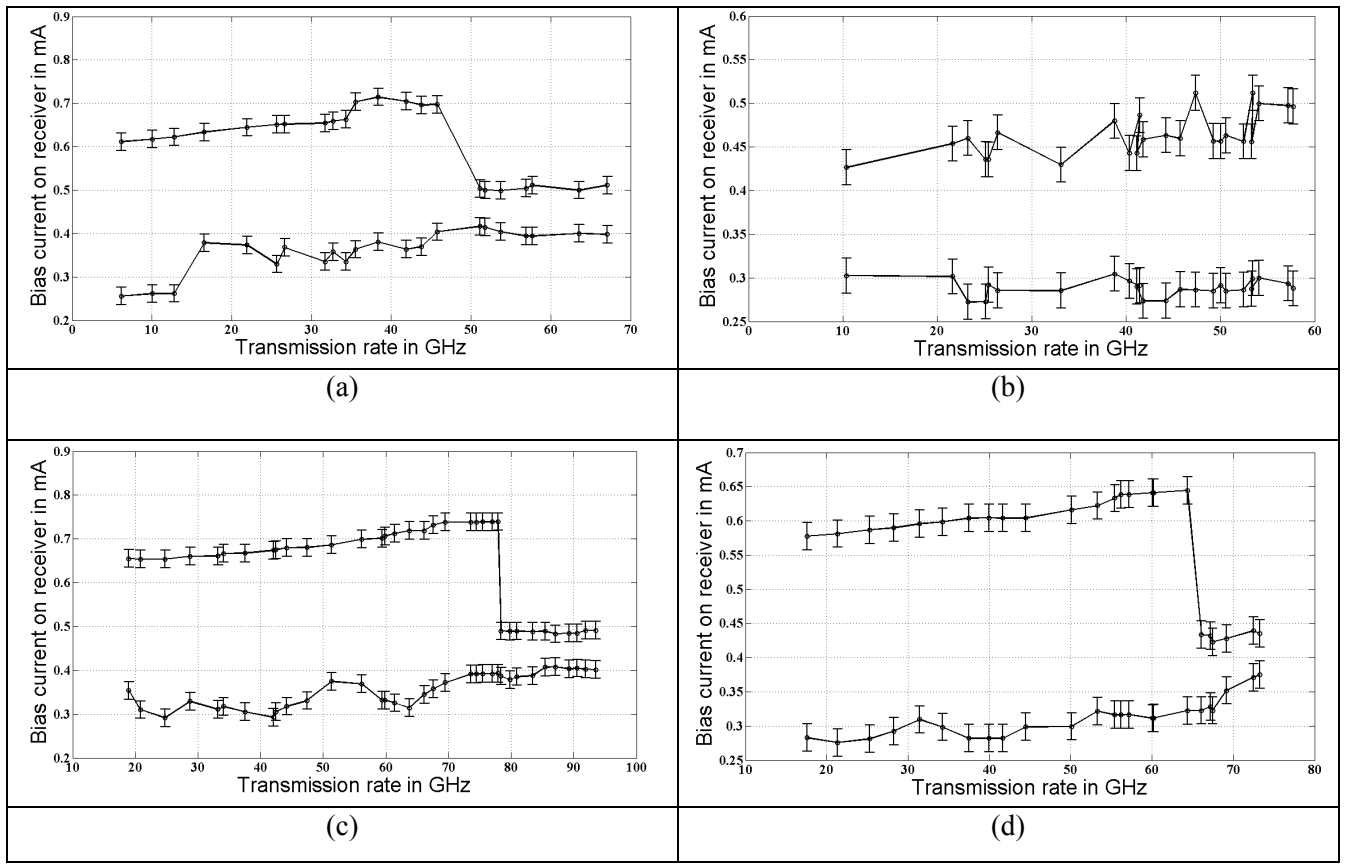

Figure 9. The upper and lower margins of the receiver bias current as a function of SFQ pulse transmission rate between the flip-chip and the MCM carrier for test circuits with different bump pad designs as indicated in Table 1: (a) Expt. 6; (b) Expt. 4; (c) Expt. 9; (d) Expt. 10

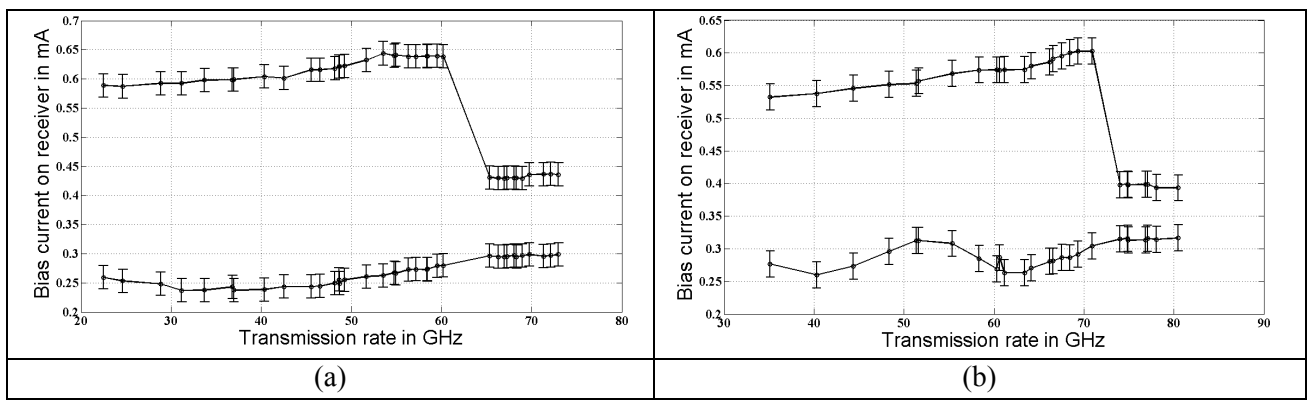

Figure 10. The upper and lower margins of the receiver bias current as a function of SFQ pulse transmission rate between the flip-chip and the MCM carrier for test circuits with different bump pad designs as indicated in Table 1: (a) Expt. 3; (b) Expt. 9. These test circuits were from the second MCM assembled from chips fabricated on a different wafer.

Some of the experiments were repeated for the second MCM assembled from chips fabricated on a different wafer in order to check the reproducibility of the results. Figure 10 illustrates the results of two measurements (Expt. 3 and Expt. 9), demonstrating quite low spread of performances between different MCMs. We can see that the margins of the receiver are similar to the shown in Figure 8(b) and Figure 9(c) data for the first MCM. The reproducibility of performance of the test circuits is important in establishing the reliability of the interconnect design and MCM assembly yield.

## 6. Conclusion

In this paper, we described and experimentally proved our design technique for MCM interconnects with ultra-high chip-to-chip data transfer rates in flip-chip MCMs. We carried also some stress tests. In particular, we allowed the microstrip lines to pass through several (up to 4) bump bonds and intentionally distorted bump pad geometry away from the optimal. All stressed circuits remained operational, and even the lowest observed chip-to-chip transfer rate (60 GHz) well exceeds (a factor of 2x) the maximum clock rates achievable in simultaneously fabricated complex digital circuits using HYPRES 4.5 kA/cm$^2$ fabrication technology. This means that the developed interconnect design, wafer bumping and flip-chip MCM assembly are fully suitable for this as well as the next node of the superconductor integrated circuit fabrication technology where the clock rates are expected to rise to ~ 60 GHz. Bump-bonded chip-to-chip connections remained stable within the entire investigated range (1.8 mm to 20 mm) of MSL lengths. Some MSLs contained up to twenty 90-degree turns. We did not detect any problems with magnetic flux that could be trapped in the bump pads. The measured MCMs successfully passed mechanical and thermal stressing, going over 100 thermo-cycles. All these observations could be used as a proof that the developed design technique is quite robust against unavoidable deviations in chip fabrication and MCM assembly technologies and delivers about 100 GHz data transfer rate at transferring pico-second SFQ pulses between integrated circuits.

## 7. Acknowledgment

This work was supported in part by ONR under contract N0001409C0687. The authors would like to thank Deborah VanVechten for her interest and support. We would also like to thank Dave Donnelly for his help with construction of electroplating set-up, John Vivalda for taking SEM pictures of plated bumps, T. Filippov for discussions of chip design issues, and to members of HYPRES circuit fabrication team, Daniel Yohannes, Rick Hunt, and John Vivalda, for their contribution to chip fabrication.

## References

1. R. L. Kautz, "Miniaturization of normal-state and superconducting microstrip lines," *J. Res. NBS,* vol. 84*,* pp. 247-259, Feb. 1979.
2. H.C. Jones and D.J. Herrell, "The characteristics of chip-to-chip signal propagation in a package suitable for superconducting circuits*," IBM J. Res. Develop.* vol. 24, pp. 172-7, March 1980.
3. S. V. Polonsky, V. K. Semenov, and D. F. Schneider, "Transmission of single-flux-quantum pulses along superconducting microstrip lines," *IEEE Trans. Appl. Supercond.* vol. 3, pp. 2598-2601, 1993.
4. Y. Hashimoto, S. Yorozu, Y. Kameda, and V.K. Semenov, "A design approach to passive interconnects for single flux quantum logic circuits," *IEEE Trans. Appl. Supercond.* vol. 13, pp. 535-38, June 2003.
5. L. A. Abelson and G. L. Kerber, "Superconductor integrated circuit fabrication technology", *IEEE Proceedings*, vol. 92, p. 1517, 2004.

6. D. Yohannes, S. Sarwana, S.K. Tolpygo, A. Sahu, Y.A. Polyakov, and V.K. Semenov, "Characterization of HYPRES' 4.5 kA/cm$^2$ and 8 kA/cm$^2$ Nb/AlO$_x$/Nb fabrication processes," *IEEE Trans. Appl. Supercond.* vol. 15, pp. 90-93, June 2005.
7. S. K. Tolpygo, D. Yohannes, R. T. Hunt, J. A. Vivalda, D. Donnelly, D. Amparo, and A. F. Kirichenko, "20 kA/cm$^2$ process development for superconductor integrated circuits with 80 GHz clock frequency," *IEEE Trans. Appl. Supercond.* vol. 17, pp. 946–951, June 2007.
8. N. Joukov, Y. Hashimoto, V. Semenov, "Matching Josephson junctions with microstrip lines for SFQ pulses and weak signals", *IEICE Trans. on Electron.*, vol. E85C, #3, pp. 636-40, Mar. 2002.
9. Q. P. Herr, A. D. Smith, and M. S. Wire, "High speed data link between digital superconductor chips," *Appl. Phys. Lett.* vol. 80, p. 32310-12, 2002.
10. P.E. Garrou and I. Turlik, "Multichip module technology handbook," McGraw-Hill, 688 pages, 1998.
11. Q. Herr, M. Wire and A. Smith, "Ballistic SFQ signal propagation on-chip and chip-to-chip", *IEEE Trans. Appl. Supercond.* vol. 13, pp. 463-466, June 2003.
12. Y. Hashimoto, S. Yorozu, and T. Miyazaki, "Transmission of single-flux-quantum pulse between superconductor chips", *Appl. Phys. Lett.* vol. 86, pp. 072502-1, 2005.
13. Y. Hashimoto, S. Yorozu, T. Satoh, and T. Miyazaki, "Demonstration of chip-to-chip transmission of single-flux-quantum pulses at throughputs beyond 100 Gbps," *Appl. Phys. Lett.* vol. 87, p. 022502, 2005.
14. Y. Hashimoto, S. Yorozu, and Y. Kameda, "Development of cryopackaging and I/O technologies for high-speed superconductive digital systems," *IEICE Trans. Electron.*, vol. E91-C, p. 325-332, Mar. 2008.
15. M.R. Rafique, H. Engseth, and A. Kidiyarova-Shevchenko, "Optimization of high frequency flip-chip interconnects for digital superconducting circuits," *Supercond. Sci. Technol.*, vol. 19, S354-S361, 2006.
16. S. K. Tolpygo, D. Tolpygo, R.T Hunt, S. Narayana, Y.A. Polyakov and V. K. Semenov, "Wafer bumping process and inter-chip connections for ultra-high data transfer rates in multi-chip modules with superconductor integrated circuits," *IEEE Trans. Appl. Supercond.*, vol. 19, No. 3, pp. 598-602, June 2009.
17. URL: www.sonnetsoftware.com/products/sonnet-suites
18. HYPRES Design Rules. URL: http://www.hypres.com/foundry/niobium-process/
19. "A guide to indium plating," Indium Corporation of America. URL: www.indium.com
20. S. B. Kaplan, V. Dotsenko, and D. Tolpygo, "High-speed experimental results for an adhesive-bonded superconducting multi-chip module," *IEEE Trans. Appl. Supercond.* vol. 17, p. 971-74, June 2007.
21. D. Gupta, T. V. Filippov, A. F. Kirichenko, D. E. Kirichenko, I. V. Vernik, A. Sahu, S. Sarwana, P. Shevchenko, A. Talalaevskii, and O. A. Mukhanov, "Digital channelizing radio frequency receiver," *IEEE Trans. Appl. Supercond.*, vol. 17, pp. 430-437, Jun. 2007.
22. V. Vernik, D. E. Kirichenko, V. V. Dotsenko, R. Miller, R. J. Webber, P. Shevchenko, A. Talalaevskii, D. Gupta, and O. A. Mukhanov, "Cryocooled wideband digital channelizing RF receiver based on low-pass ADC," *Supercond. Sci. Technol.* vol. 20, pp. S323-S327, Nov. 2007.
23. R. J. Webber, V. Dotsenko, A. Talalaevskii, R. Miller, J. Tang, D. E. Kirichenko, I. V. Vernik, P. Shevchenko, V. Borzenets, D. Gupta, and O. A. Mukhanov, "Operation of superconducting digital receiver circuits on 2-stage Gifford-McMahon cryocooler," *Advances in Cryogenic Engineering:*

*Transactions of the Cryogenic Engineering Conference (CEC 2007)*, vol. 53; *AIP Conference Procs.* vol. 985, pp. 927-932, Mar. 2008.
24. O. A. Mukhanov, D. Kirichenko, I. V. Vernik, T. V. Filippov, A. Kirichenko, R. Webber, V. Dotsenko, A. Talalaevskii, J. C. Tang, A. Sahu, P. Shevchenko, R. Miller, S. B. Kaplan, S. Sarwana, and D. Gupta, "Superconductor Digital-RF receiver systems," *IEICE Trans. Electron.*, vol. E91-C, No. 3, pp. 306-317, Mar. 2008.
25. D. Gupta, D. E. Kirichenko, V. V. Dotsenko, R. Miller, S. Sarwana, A. Talalaevskii, J. Delmas, R. J. Webber, S. Govorkov, A. F. Kirichenko, I. V. Vernik, and J. Tang, "Modular, multi-function digital-RF receiver systems," *IEEE Trans. Appl. Supercond.* vol. 21, pp. 883-890, 2011.
26. J. S.E. Ranjith, C.S. Thompson, V.V. Dotsenko, J. Delmas, A.P. Malshe, D. Gupta, "Carbon nanotube based polymer adhesive as an underfill for superconductor multi-chip module packaging," *IEEE Trans. Appl. Supercond.* vol. 21, No. 3, pp. 900-03, June 2011.
27. S. Narayana, Y.A. Polyakov, V.K. Semenov, "Evaluation of Flux Trapping in Superconducting Circuits," *IEEE Trans. Appl. Supercond.* vol.19, No.3, pp.640-43, June 2009.
28. R.P. Robertazzi, I. Siddiqi, O. Mukhanov, "Flux trapping experiments in single flux quantum shift registers," *IEEE Trans. Appl. Supercond.*, vol. 7, No. 2, pp. 3164–7, 1997.